# The detection of nanoscale membrane bending with polarized localization microscopy

A. M. Kabbani and C. V. Kelly

## ABSTRACT

The curvature of biological membranes at the nanometer scale is critically important for vesicle trafficking, organelle morphology, and disease propagation. The initial membrane bending events that initiate these complex processes occur at a length scale that is irresolvable by most super-resolution optical microscopy methods. This manuscript reports the development of polarized localization microscopy (PLM), a pointillist optical imaging technique for the detection of nanoscale membrane curvature in correlation with single-molecule dynamics and molecular sorting. PLM combines polarized total internal reflection fluorescence microscopy (TIRFM) and single-molecule localization microscopy to reveal membrane orientation with a sub-diffraction-limited resolution without reducing localization precision by point spread function (PSF) manipulation. Membrane curvature detection with PLM requires fewer localization events to detect curvature than 3D single-molecule localization microscopy (*e.g.*, PALM or STORM), which enables curvature detection 10x faster via PLM. With rotationally confined lipophilic fluorophores and the polarized incident fluorescence excitation, membrane-bending events are revealed with super-resolution. Engineered hemispherical membrane curvature with a radii ≥ 24 nm was detected with PLM and individual fluorophore localization precision was 13 ± 5 nm. Further, deciphering molecular mobility as a function of membrane topology was enabled. The diffusion coefficient of individual DiI molecules was 7.6x higher in planar supported lipid bilayers than within nanoscale membrane curvature. Through the theoretical foundation and experimental demonstration provided here, PLM is poised to become a powerful technique for revealing the underlying biophysical mechanisms of membrane bending at physiological length scales.

**INTRODUCTION**

Nanoscale membrane curvature is essential for many biological functions (1), including the regulation of lipid rafts (2), exocytosis/endocytosis (3), viral fusion/egress (4), nano-therapeutics (5), membrane remodeling (6), and the shedding of circulating microvesicles (7). Membrane curvature can be induced by the line tension between coexisting liquid lipid phases, the aggregation of curvature preferring molecules, the steric pressure between crowded proteins, and the molecular shape of either lipids or proteins (8–10). However, quantifying the relative contributions of these curvature-generating mechanisms at physiological length scales remains elusive due to limited experimental capabilities to detect nanoscale bending. In this manuscript, we report the development of polarized localization microscopy (PLM), which combines single-molecule localization microscopy (SMLM) with polarized total internal reflection fluorescence microscopy (TIRFM). Polarized TIRFM distinguishes between membranes (11) and molecules (12) of varying orientation by measuring the overlap between the fluorophore's transition dipole moment and linearly polarized incident excitation light. Indocarbocyanine dyes (*e.g.*, DiI) maintain their transition dipole moment in the plane of the membrane such that DiI in membranes parallel to the coverslip is preferentially excited by incident s-polarized light and DiI in membranes vertical to the coverslip are preferentially excited by incident p-polarized light (Fig. 1)(13–15). Diffraction-limited polarized TIRFM has advanced the detection of membrane curvature despite lateral resolution limited to >200 nm (16), as demonstrated with presynaptic vesicle fusion (17) and endocytosis/exocytosis (11, 18).

Super-resolution SMLM such as fluorescence photo-activated localization microscopy ((f)PALM) and direct stochastic optical reconstruction microscopy ((d)STORM) have overcome the diffraction-limited resolution of traditional optical microscopy to provide images with a lateral resolution of <20 nm (19–22). SMLM is the computational localization of individual fluorophores that are sparsely blinking in sequential diffraction-limited images for the reconstruction of a super-resolution image. The resolution of the resulting reconstructed image depends on localization imprecision, systematic inaccuracies, and limited sampling densities (23, 24). 3D SMLM has been implemented through a variety of methods, including the insertion of a cylindrical lens into the light path (25), single-fluorophore interference in a 4π configuration (26), biplane imaging[26], and emission phase manipulations (28). For these methods, information about the fluorophore vertical location requires either incorporating a precisely aligned multi-camera interferometric detection path or sacrificing precision in lateral localization through the manipulation of the point spread function (PSF) to yield localization precisions along the *z*-direction ($\sigma_z$) typically double that of the *xy*-plane ($\sigma_{xy}$), approximately equal to 40 nm and 20 nm, respectively. Similarly, single-fluorophore orientations have been measured precisely by combinations of image defocusing, emission phase modulations, steerable filters, and advanced fitting routines (28–36). However, many of these methods are optically and computationally challenging in addition to the complication of multicolor imaging and requiring hundreds of localizations to identify nanoscale membrane buds.

PLM provides fluorophore orientation with conventional detection optics, no sacrifice of localization precision by PSF manipulation, minimal adjustment of the excitation optics, and the use of commercial fluorophores. TIRF SMLM setups are able to perform PLM by the sole addition of a liquid crystal variable wave plate (*i.e.,* Thorlabs Inc, LCC1111-A), which controls the polarization of the fluorescence excitation based on the voltage applied to it. PLM depends on the use of rotationally confined fluorophores that maintain an orientation relative to the membrane normal (37) and photoswitch between fluorescent bright and transient dark states (38, 39), such as the indocarbocyanine dye DiI. Super-resolution images from PLM are reconstructed such that membranes vertical to the coverslip primarily yielded localizations from p-polarized localization microscopy (pPLM) and membranes parallel to the coverslip was best observed in s-polarized localization microscopy (sPLM). This approach enables resolving dynamic nanoscale membrane curvature under aqueous, physiological conditions to reveal curvature-induced variations in membrane organization and dynamics.

In this manuscript, PLM was used to detect engineered nanoscale membrane curvature and correlate curvature to single-molecule trajectories. The capabilities of PLM were theoretically calculated for varying membrane topologies in support of the order-of-magnitude improvements in detection and resolution of membrane curvature (Fig. 2). Curvature in model membranes was created by draping supported lipid bilayers (SLBs) over nanoparticles (NPs) of known sizes, ranging in radius ($r_{NP}$) from 24 to 70 nm. The resulting membrane curvature and curvature-influenced diffusion of individual lipids were resolved. PLM provided detection and visualization of nanoscale curvature in agreement with theoretical predictions. In sum, these studies demonstrate the capabilities of PLM to advance optical imaging capacities while providing order-of-magnitude improvements in spatial and temporal resolution than comparable SMLM techniques.

## MATERIALS AND METHODS

### Sample dish preparation

Glass bottom dishes (MatTek Corp.) were cleaned by immersion in 7x detergent overnight, rinsed with diH2O, bath sonicated for 30 min, dried with nitrogen gas, and cleaned by air plasma (Harrick Plasma). NPs were diluted in diH$_2$O, sonicated for 15 min, and deposited on a glass coverslip. NP sedimentation occurred for 10 min to achieve a density of 0.02 NPs/µm$^2$. Separate polystyrene NPs were used for both engineering membrane curvature and tracking stage drive. The index of refraction of polystyrene is 1.59. NPs for creating membrane curvature were either 29 nm radius, $\lambda_{ex}$ = 647 nm (FluoSpheres, Life Technologies); 51 nm radius, $\lambda_{ex}$ = 405 nm (FluoSpheres, Life Technologies); or 70 nm radius, $\lambda_{ex}$ =488 nm (Fluoro-Max, Fisher Scientific). NPs for detecting stage drift (100 nm diameter, Tetraspecs, Life Technologies) were fluorescent in all color channels. Dishes were placed on a 55 °C hot plate for 5 min to ensure their stability on the coverslip.

### Supported lipid bilayer formation

Giant unilamellar vesicles (GUVs) of primarily 1-palmitoyl-2-oleoyl-sn-glycero-3-phosphocholine (POPC, Avanti Polar Lipids, Inc.) labeled with 0.3 mol% 1,1'-didodecyl-3,3,3',3'-tetramethylindocarbocyanine perchlorate (DiI, Life Technologies) were prepared by electro-formation, as described previously[51]. Details on the GUV formation method are provided in the supplemental material. This fluorophore density yielded 110 nm$^2$ of bilayer per DiI molecule. The interaction between the GUVs with the plasma cleaned glass coverslip resulted in bursting of the GUVs and the formation of patches of SLB over the glass and NPs. This method of SLB creation proved to create more uniform SLBs over the NPs than SLBs formed by the fusion of large unilamellar vesicles (LUVs). The detailed methods for LUV creation are provided in the Supplemental Material.

### Optical setup

PLM was performed with an inverted IX83 microscope with Zero-Drift Correction and a 100x, 1.49NA objective (Olympus Corp.) on a vibration-isolated optical table. Four continuous wave diode lasers were incorporated at wavelengths 405, 488, 561, and 647 nm, each with at least 120 mW max power for fluorescence excitation. The excitation polarization was rotated with computer-controlled liquid crystal waveplate (Thorlabs Inc, LCC1111-A). Image acquisition was performed with an iXon-897 Ultra EMCCD camera (Andor Technology) proceeded by emission filters (BrightLine single-band bandpass filters, Semrock, Inc.), a 4-band notch filter (ZET405/488/561/640m, Chroma Corp.), and a 2.5x magnification lens (Olympus Corp). This setup provided high laser power (>80 mW) at each polarization and integrated computer control of all equipment via custom LabVIEW routines (National Instruments Corp.).

**Imaging procedure**

Exposure of the sample to >80 mW of excitation light with $\lambda_{ex}$ = 561 nm for 3 s resulted in converting most of the DiI from the fluorescent state '*on*' to the transient non-fluorescent, dark state '*off*' to provide steady state fluorophore blinking. The '*on*' fluorophores were imaged at a density of less than one fluorophore per 1 μm$^2$. Sequential movies were acquired with alternating p-polarized TIRF (pTIRF) excitation at $\lambda_{ex}$ = 561 nm for pPLM and s-polarized TIRF (sTIRF) excitation at $\lambda_{ex}$ = 561 nm for sPLM. 10,000 to 30,000 frames were acquired for each polarization at a frame rate of 50 Hz on a region of interest with 18 ms acquisition per frame.

**Imaging buffer**

PLM was performed on samples present in an oxygen-scavenging buffer (150 mM NaCl, 50 mM TRIS, 0.5 mg/mL glucose oxidase, 20 mg/mL glucose, and 40 μg/mL catalase at pH 8). Buffer proteins were purchased from Sigma-Aldrich and salts were purchased from Fisher Scientific. These conditions maintain a low free oxygen concentration in the buffer to minimize non-reversible fluorophore bleaching and encourage transient fluorophore blinking, as is necessary for SMLM.

**Single-fluorophore localizations**

The analysis of the raw, diffraction-limited images included low-pass Gaussian filtering, median background subtraction, lateral stage drift correction, and the fitting of each isolated fluorophore images via the ImageJ plug-in ThunderSTORM (40). ThunderSTORM provided the single fluorophore positions, localization uncertainty, and photon per fluorophores for further analysis. A threshold value 100 photons per fluorophores was used to keep only the bright localizations for further analysis. Single-molecule DiI localizations had 13 ± 5 nm precision (Table S1, Fig. S1). The localizations from s- and pTIRF excitation were analyzed separately to reconstruct separate super-resolution images for each membrane orientation.

**Single-particle tracking (SPT)**

The sequential localizations of single fluorophores were analyzed to reveal the diffusion rate of individual molecules versus membrane topography. The individual fluorophore trajectories projected onto the imaging plane were identified with custom MATLAB code. Single-fluorophore localizations were linked as a trajectory if they were in sequential frames, within 500 nm of each other, and there was no alternative localization for linking within 1 μm. The link lengths ($v$) were grouped based on their distance from the NP center, and their normalized distribution was fit to a 2D Maxwell-Boltzmann distribution (Eq. 1) as would be expected for 2D Brownian diffusion.

$$P(v) = \frac{v}{2D'_{xy}\Delta t} e^{-\frac{v^2}{4D'_{xy}\Delta t}} \tag{Eq. 1}$$

The projection of the lipid trajectories onto the imaging plane yielded a decrease in their apparent step lengths depending on the membrane tilt ($\theta$); this effect is considered in the simulations of single molecule trajectories described below. The localization imprecision ($\sigma_{xy}$ = 13 ± 5 nm) increased the apparent step lengths. Accordingly, the reported diffusion coefficient through the $xy$-plane ($D_{xy}$) was calculated from the fit of Eq. 1 according to $D_{xy} = D'_{xy} - \sigma^2/2/\Delta t$.

Whereas diffusion coefficients are typically extracted from a trajectory by fitting the mean squared displacement versus $\Delta t$ for $\Delta t > 0$, this routine was not appropriate here. Fitting a longer trajectory to a single diffusion coefficient would have blurred the effects of curvature with the lipid trajectory sampling both curved and flat membranes, as discussed below.

**Data analysis calculations**

Signal-to-noise calculations of diffraction-limited images were performed by taking the ratio of the mean intensity difference at the membrane bud divided by the standard deviation of the intensity of the surrounding planar SLB. Whereas, the signal-to-noise ratio (SNR) for the super-resolution reconstructed images was evaluated through dividing the mean signal, calculated from the number of localizations at the curvature location, by the standard deviation of the number of localizations of the flat bilayer.

The size of each membrane bud ($<r>$) was set equal to the mean distance from the bud center of all extra localizations due to the bud. This was calculated by taking into consideration the background from flat SLB localizations of uniform density ($\rho$), the distance of each localization from the bud center ($r_i$), and a threshold distance that was significantly greater than $<r>$ ($R$). Typically, $R = 400$ nm but the following calculation is independent of the particular $R$ chosen. The number of extra localizations due to the presence of the bud ($N_{bud}$) is equal to the total number of localizations ($N_{all}$) within $r_i < R$ subtracted from the number of localizations expected within $R$ if no bud was present ($N_{SLB}$); $N_{SLB} = \pi R^2 \rho = N_{all} - N_{bud}$. The mean $r_i$ expected for the flat SLB within $R$ is $2R/3$. By analyzing all collected localizations within $R$ and subtracting the expected localizations from the flat SLB, $<r>$ is calculated according to

$$<r> = \frac{\sum r_i}{N_{bud}} - \frac{2\pi \rho R^3}{3 N_{bud}}. \tag{Eq. 2}$$

**Modeled membrane topography and diffusion**

Simulated membrane topography was created by smoothly connecting a spherical NP coating to a planar sheet with no less than 20 nm radii of curvature (Fig. 5C). The larger NPs formed a neck-like structure of radius smaller than the NPs, and the smaller NPs form a tent-like structure extending beyond the NP radii. With custom MATLAB routines, a random distribution of points on these simulated topographies mimicked the possible 3D locations of localized fluorescent lipids. These points were used to reconstruct simulated PLM images and lipid trajectories by incorporating the localization probabilities discussed below.

Broadly, single-fluorophore images may laterally shift if the fluorophores possess limited rotational freedom and anisotropic fluorescence emission. The effects of rotationally confined fluorophores can yield lateral localization inaccuracies up to 100 nm upon defocusing by 200 nm (41). Numerical integration yielded the magnitude and direction of the shift in localization position due to the single fluorophore orientation and height above the focal plane following the framework of Agrawal et al. (42). The expected PSF and lateral shift were estimated as a function of membrane orientation ($\theta$ and $\varphi$) by averaging together the expected fluorophore orientations with the membrane ($\psi$ and $\beta$). Accordingly, the expected lateral shifts as a function of membrane orientation and height were calculated. This systematic shift was incorporated into our simulated image reconstruction and SPT results, proving to be critical for matching the experimental data. Since the magnitude of the anisotropic emission effects vary greatly with distance between the single fluorophore in the membrane and the focal plane, and this distance was difficult to experimentally assess, the magnitude of defocusing and lateral shifting was fit to match experimental and theoretical results.

## RESULTS

### Theory of PLM

The number of localizations to be collected from various membrane topologies by PLM was theoretically estimated. PLM depends on the relative orientation between the DiI fluorescence dipole moment ($\mu$) with the fluorescence excitation light ($E$). The coordinate frame of the microscope is defined such that the coverslip-water interface is in the *xy*-plane with $z = 0$. The local membrane orientation is represented by the polar ($\theta$) and azimuthal ($\varphi$) angles of the membrane normal vector relative to the microscope coordinate frame. Relative to the membrane normal, the DiI fluorescence dipole moment experiences a polar tilt ($\beta$) and azimuthal rotation ($\psi$). Therefore, the Cartesian components of $\mu$ are

$$\mu_x = \cos\theta \cos\phi \sin\beta \cos\psi - \sin\phi \sin\beta \sin\psi + \sin\theta \cos\phi \cos\beta,$$
$$\mu_y = \cos\theta \sin\phi \sin\beta \cos\psi + \cos\phi \sin\beta \sin\psi + \sin\theta \sin\phi \cos\beta,$$
$$\mu_z = \cos\theta \cos\beta - \sin\theta \sin\beta \cos\psi, \tag{Eq. 3}$$

as shown previously(14). The polar tilt of DiI in the membrane has been previously measured to be $\beta = 69°$. Changing $\beta$ by 5° has a <5% effect on these results(18). The azimuthal rotation of DiI samples all angles within 0.2 ns(14), resulting in an averaging over $\psi$ for the 100s of excitation events that occur during the 18 ms single-frame exposure time. Further, the high numerical objective used in these experiments (NA = 1.49) yields collection efficiency consistent within 10% for the emission of all fluorophore orientations(18).

Within our experimental setup, the p-polarized evanescent field ($E_p$) is elliptically polarized in the *x-z* plane and the s-polarized evanescent field ($E_s$) is linearly polarized in the *y*-plane according to(14, 43)

$$\boldsymbol{E}_p = E_p^o(0.5\boldsymbol{x} + 1.9i\boldsymbol{z}) \cdot \exp\left(\frac{-z}{2d}\right),$$

$$\boldsymbol{E}_s = E_s^o(1.7\boldsymbol{y}) \cdot \exp\left(\frac{-z}{2d}\right), \quad \text{(Eq. 4)}$$

where $E_p^o$ and $E_s^o$ represent the magnitude of the p- and s-polarized incident electric field, respectively(43). The penetration depth of the evanescent field ($d$) was 124 nm, as determined by the excitation incident angle ($\theta_i = 65°$), excitation wavelength ($\lambda_{ex} = 561$ nm), and the indices of refraction of the sample and glass (1.33 and 1.52, respectively). Approximating $\boldsymbol{E}_p$ to have a zero $x$ component induces 7% error and simplifies the intensity of excitation for each DiI molecule as a function of the membrane orientation equal to $I_p = (\mu_z \cdot \boldsymbol{E}_p)^2$ and $I_s = (\mu_y \cdot \boldsymbol{E}_s)^2$ for p- and s-polarized excitation, respectively.

Diffraction-limited polarized TIRFM compares $I_p$ and $I_s$ directly. PLM incorporates these intensities into the localization probability and provides increased sensitivity to changes in fluorophore orientation. Individual fluorophores demonstrated an exponential distribution of detected brightness with the average fluorophore brightness proportional to $I_p$ or $I_s$ (Fig. S1). Only fluorophores with a detected brightness greater than a brightness threshold ($B_0$) were localized for inclusion in the PLM results. Therefore, the probability of detecting a DiI molecule as a function of membrane orientation ($\theta$, $\varphi$), DiI orientation with the membrane ($\beta$, $\psi$), and height above the coverslip ($z$) was approximated as

$$P_p = \exp\left(\frac{-B_0}{3.6\langle\mu_z^2\rangle\exp(-z/d)}\right),$$

$$P_s = \exp\left(\frac{-B_0}{2.9\langle\mu_y^2\rangle\exp(-z/d)}\right), \quad \text{(Eq. 5)}$$

for p- and s-polarized excitation, respectively, where < > represents the average over all $\psi$. $B_0$ was set to match these theoretical results to the experimental results (Fig. 5). Increasing the brightness threshold increase pPLM sensitivity to $\theta$. $P_s$ and $P_p$ were compared to the expected detection probability for hypothetical unpolarized excitation light ($P_{uTIR}$), for which all fluorophore orientations had a 50% overlap between the fluorescent dipole moment and exciting electric field direction such that there is no $\theta$, $\varphi$, $\beta$, or $\psi$ dependence on $P_{uTIR}$,

$$P_{uTIR} = \exp\left(\frac{-B_0}{1.7\exp(-z/d)}\right). \tag{Eq. 6}$$

The probability of detecting a fluorophore with unpolarized epifluorescence illumination would have no $\theta$, $\varphi$, $\beta$, $\psi$, or $z$ dependence and it would represent the probability of detecting fluorophores with unrestricted orientations, such as fluorophores bound by flexible linkers. In a well-sampled image, pPLM yields the magnitude of $\theta$ and no information on $\varphi$, and sPLM results depend on both $\theta$ and $\varphi$. Neither the sign of $\theta$ nor the value of $\varphi$ may be determined with only the $z$-polarized and $y$-polarized excitations; resolving changes to $\theta$ across a sample is sufficient for detecting membrane curvature.

**Comparison between SMLM methods**

Membrane topologies were simulated with these detection probabilities to demonstrate the effects of polarization and total internal reflection (TIR) in super-resolution imaging. The sensitivity of pPLM, sPLM, unpolarized TIR-SMLM, unpolarized Epi-SMLM, and unpolarized 3D-TIR-SMLM to membrane budding was calculated. A 50 nm radius membrane vesicle was simulated budding from a planar SLB (Fig. 2A), the expected number of localizations increases for all polarizations since the area of the membrane and the number of fluorophores increase at the site of budding. Illumination modes were compared by predicting the increase in the expected number of localizations for vesicle budding (Fig. 2B). The number of localizations from p-polarized excitation increased more than other polarizations because the new membrane had large $\theta$. Upon the formation of a hemispherical membrane bud, when the top of the bud was 50 nm above the surrounding SLB, 11x more localizations would be detected by pPLM because of the presence of the bud. When the bud top is 100 nm away from the SLB and a full formed vesicle, pPLM would yield 23x more localizations while other illumination methods reveal at most 5x more localizations (Fig. 2B).

3D SMLM is able to reveal membrane bending from the height of each localization rather than the change in a number of localizations. However, the uncertainty of each fluorophores height requires the averaging of multiple fluorophore's height in a given region to confidently detect the membrane height. Due to this uncertainty in the fluorophore height, using the height from 3D SMLM requires more localizations than analyzing the differences in local density of localizations obtained from the extra membrane area associated with the bud for 50 nm radius buds that are within 90 nm from the surrounding SLB.

The number of localizations in a unit area is expected to obey a Poisson statistics. A membrane bud is identified with statistical significance of given p-value (*p*) when the number of localizations at the membrane bud ($N_{bud}$) is greater than the average number of localizations over the planar membrane ($N_{plane}$) according to

$$p = e^{-N_{plane}} \sum_{i=N_{bud}}^{\infty} \frac{(N_{plane})^i}{i!}.$$ (Eq. 7)

The minimum value of $N_{bud}$ that satisfies this equation ($N_{bud}^0$) varies with the ratio of $N_{bud}$ to $N_{plane}$, which depends on the membrane topography and excitation polarization. For instance, for a polarization sensitive dye (*e.g.*, DiI) and p-polarized excitation, $N_{bud}$ can be over 20x greater than $N_{plane}$ (Fig. 2B).

3D SMLM provides a minimal advantage in the ratio of $N_{bud}$ to $N_{plane}$, but it does provide the height of each fluorophore with an associated uncertainty ($\sigma_z \approx 40$ nm). For 3D SMLM, statistically significant bud detection occurs when the average *z*-value of the localizations at the bud ($<z_{bud}>$) has a standard error of the mean (SEM) sufficiently small such that the bud may be distinguished from the surrounding planar membrane at *z* = 0, according to

$$p = 0.5 \, erfc\left( \langle z_{bud} \rangle \frac{\sqrt{N_{bud}^0}}{\sigma_z \cdot \sqrt{2}} \right),$$ (Eq. 8)

where *erfc* is the complementary error function. The integer $N_{bud}^0$ that satisfies Eqs. 7 and 8 for pPLM and 3D SMLM, respectively, were calculated and plotted (Fig. 2C). pPLM is predicted to provide statistically significant membrane bud identification with fewer localizations than 3D SMLM for all 50 nm diameter buds. Only after the bud has undergone vesicle fission and the top of the bud is 110 nm away from the SLB does 3D SMLM require fewer localizations to detect the bud than pPLM.

For instance, PLM requires only 10% of the total number of localizations required for 3D SMLM with *p* = 0.0001 when the bud top is 50 nm above the surrounding planar membrane. This corresponds to detecting membrane budding 10x faster via pPLM than 3D SMLM. Membrane curvature detection with 3D SMLM requires data averaging that further reduces sensitivity and resolution, whereas pPLM localization density itself is correlated with membrane bending. Experimental evidence comparing localization rate and time required to detect local membrane bending is later discussed in reference to Fig. 6.

The relatively small change in the number of localizations that are detected by sPLM upon membrane budding (<3.5x) provides an internal control for other possible membrane topographies. Whereas pPLM may yield a significant increase in localizations due to the bud, sPLM shows apparently random fluctuations across the sample. A high local density of pPLM localizations is more confidently considered to occur due to membrane bending when coincident with a nearly uniform distribution of sPLM localizations. This case is opposed to localization densities caused, for instance, by a chromatic bleed through from another fluorescence source, that would cause an increase in localizations in both pPLM and sPLM.

**Resolution and sensitivity of PLM**

To demonstrate the ability of PLM to detect membrane curvature, we created membrane bending by three different methods: SLBs draped over NPs (Figs. 3, S3, and S4); LUVs above an SLB (Fig. S2); and unfused GUVs adhered to the glass coverslip (Fig. S5). The best control and consistency of the engineered membrane curvature came from the SLBs draped over the NPs. SLBs composed of 99.7 mol% POPC and 0.3 mol% DiI were draped over NPs to create the desired membrane topography. This procedure was reproduced for hundreds of NPs of $r_{NP}$ = 24, 51, and 70 nm. In all cases, pPLM provided an increased density of localizations at the site of membrane curvature (Fig. 4) and chromatic bleed through from the NP was not present (Fig. S7). For example, the density of localizations at the curved membrane over the 70 nm NP in pPLM was (2.2 ± 1) x$10^{-6}$ localizations/nm$^2$/frame, a 27x increase over (8.2 ± 3) x$10^{-8}$ for flat SLB (Fig. 5). As an important internal control, no significant increase in the number of sPLM localizations was observed with nanoparticle-induce membrane curvature (Fig. 3J).

Comparisons between the diffraction-limited images of the fluorescent polystyrene NP, the diffraction-limited polarized TIRFM images, and the reconstructed PLM images of the membrane reveals the increased resolution and detection sensitivity provided by PLM (Fig. 3). Tthe diffraction-limited images demonstrated the PSF of the microscope more so than the physical size of the nanoparticle or membrane curvature. However, the radius of each membrane bud ($\langle r \rangle$) was calculated from pPLM images by averaging the distance between each localization and the center of the bud (*r*). This calculation yielded <r> of 32 ± 4, 50 ± 14, and 60 ± 13 nm for membrane draped over NPs of 24, 51, and 70 nm radii, respectively (Fig. 4). Greater consistency in <r> calculations was provided when more localizations per area were detected (Fig. S6).

The sensitivity of PLM for detecting membrane curvature was especially apparent for the SLBs draped over NPs of 24 nm radii. The faint signal from the membrane curvature in diffraction-limited pTIRFM images could have gone undetected, whereas the increased density of localizations pPLM is readily apparent (Figs. 5 and S3). pPLM provided a 6x increase in the SNR over diffraction-limited p-polarized TIRFM with SNR of $11 \pm 9$ and $1.9 \pm 0.7$, respectively, and the uncertainty represents the standard deviation between events.

To reveal PLM temporal resolution, an autocorrelation analysis was performed on the PLM data. Correlation analysis was performed on both sPLM and pPLM images with increasing acquisition time interval to find PLM temporal resolution in detecting membrane bending events. Results reveal the increased correlation between localizations due to the curvature detection in pPLM in comparison to the more uniform localization distributions from sPLM. Localization density rate of $(1.2 \pm 0.1) \times 10^{-6}$ localizations/nm$^2$/frame, enabled early detection of local membrane bending over the 70 nm NPs within 1 sec in pPLM with p-value of 0.0239, for a 3 sec acquisition interval the curvature region is detected in pPLM with a p-value of 0.0002 (Fig. 6).

LUVs of with 0.3 mol% DiI were imaged with PLM. From diffraction-limited images of polarized TIFM excitation, flat SLBs were $(1.8 \pm 0.3)$x brighter with sTIRFM than pTIRFM with the primary variability coming from laser alignment and SLB quality. Unfused LUVs above an SLB yielded $(1.8 \pm 0.7)$ more signal from pTIRFM than sTIRFM with the major variability coming from the LUV size. The combination of these factors yielded a $3.2 \pm 0.8$ fold increase in signal for LUV detection via diffraction-limited pTIRFM versus sTIRFM. pPLM yielded a 7.6x increase in localization rate when an LUV was present over an SLB with a $(50 \pm 20)$ versus $(6.6 \pm 0.8)$ x10$^{-7}$ localizations/nm$^2$/frame in pPLM versus sPLM. The mean and standard deviation of the LUV radii was <r> $54 \pm 29$ and $57 \pm 21$ nm as measured by pPLM and sPLM, respectively As a demonstration of the increased sensitivity provided by PLM, 81% of the 122 LUVs that were detected in both sPLM and pPLM were not apparent with diffraction-limited pTIRFM or sTIRFM (Fig. S10). The LUVs only detected by PLM had radii shifted to smaller values of <r> of $62 \pm 20$ nm while LUVs detected in PLM and TIRF possessed <r> of $72 \pm 10$ nm.

Localization imprecision was limited primarily by the number of photons collected from each fluorophore in each frame. The localization software, ThunderSTORM, accounted for the camera quantum efficiency and imaging noise to estimate the number of photons and the localization precision for each detected fluorophore. 1200 ± 800 photons/fluorophore/frame were acquired yielding a localization precision of 13 ± 5 nm. Further information regarding the acquired number of photons/fluorophore and the uncertainty for the different NPs sizes are provided in the supplemental information (Fig. S1 and Table S1).

The upper limit on localization rates in all SMLM methods is based on the camera frame rate and the length scale of diffraction-limited imaging. Localization rates could be increased above those reported here by increasing density of DiI in the sample, or optimizing the DiI *on-* and *off-*rates with further buffer or incident light optimization. Further, the limited final number of localizations yields uncertainty in analyzing the precise local membrane orientation and the center of the membrane bud. Here, 200 ± 100 localizations per membrane bud were collected, each with a radius of 30 to 60 nm, which resulted in an uncertainty in identifying the curvature center by 3 ± 1 nm.

**Membrane bending affects lipid mobility**

The same raw data from PLM that reveals nanoscale membrane bending also reveals how single-lipid trajectories are affected by membrane curvature. High-throughput SPT was performed on the raw PLM data by tracking of individual fluorophores that were localized in sequential frames. Single-molecule DiI diffusion was observed with pPLM and sPLM to reveal the apparent diffusion coefficient in the *xy*-plane ($D_{xy}$). DiI that were detected in more than one frame were detected in 3.8 sequential frames on average. Analyzing $D_{xy}$ as a function of location on the sample revealed the effects of membrane topology to lipid dynamics. By analyzing $D_{xy}$ versus distance from the NP center demonstrated how the lipid diffusion slowed at the membrane buds equivalent to as if the membrane bending caused an increase in viscosity and slowed the diffusion coefficient by 7.6x versus the surrounding, planar SLB (Fig. 7).

Membrane tilt can decrease the apparent $D_{xy}$ by up to 60%; however, upon incorporating the localization imprecision, imaging frame rates, and sample averaging, no more than 45% decrease in $D_{xy}$ due to geometrical effects alone in our system. In order to reproduce our experimental data, an addition mechanisms for slowing the lipids at the sites of membrane curvature was needed.

The incorporation of a diffusion barrier that prevented single-lipid trajectories from crossing between the curved membrane bud and surrounding SLB was insufficient to reproduce the experimental data. With a 50 Hz frame rate, a local diffusion coefficient of 0.4 µm$^2$/s, and a 70 nm radius NP, a simulated diffusion barrier yielded only a 10% decrease in the observed $D_{xy}$ at the membrane bud. Further, the FRAP results demonstrate the continuity of the membrane between the bud and the SLB (Fig. S8). In sum, there is no support for a diffusion barrier limiting diffusion between the bud and the connected SLB.

The incorporation of a local lipid diffusion coefficient that was slower in curved membranes than planar membranes was able to yield an agreement between the modeled and experimentally observed the radial dependence of $D_{xy}$ (Fig. 7). Simulations of lipids diffusing on a membrane over a NP coupled to a planar membrane for which $D_{plane}$ was 7.6x greater than $D_{curved}$ reproduced the PLM experimental results for $D_{xy}$ versus distance from NP center.

## DISCUSSION

### Engineered membrane curvature

The method of creating SLBs primarily used in these studies incorporated the draping burst GUVs over the glass coverslip and polystyrene NPs. Draping a bilayer over NPs of known radii provided a model of physiologically similar dimensions to clathrin-coated pits or endocytosis/exocytosis prior to scission. SLBs created by bursting GUVs were more intact and contained fewer pores than creating a bilayer via LUV fusion (44). However, such holes within the SLB were still feasible with this GUV-fusion method, especially when GUVs were more violently ruptured via application of GUVs to freshly plasma cleaned glass coverslips or dilution with hypotonic solutions. The continuity of the membrane between the SLB and the curvature over the NP was confirmed with fluorescence recovery after photobleaching (FRAP). FRAP demonstrated that the lipids coating the NP can exchange with the lipids directly on the coverslip (Fig. S8), similar as also shown previously (45). Examination of the continuity of the bilayer over NPs was performed by assessing the pPLM data where 94% of 290 NPs surrounded by an SLB had curved membrane draping over the NP. NPs without membrane curvature could be due to the NP being on top of the bilayer rather than under it or the formation of a hole in the SLB directly surrounding the NP.

An alternative membrane topography examined is that of the LUVs bound to an underlying SLB, which resembles vesicle docking in endocytosis and the later stage of exocytosis that precedes vesicle fission in cells. The variation of LUV sizes obtained by extrusion was demonstrated by PLM. Other studies of vesicle sizes produced by extrusion have found the average diameter of extruded LUVs to be $65 \pm 30$ nm (46) indicating that the extrusion process produces a variation of LUV sizes with an upper size limit comparable to the extruder filter pore size, and a small size subpopulation of diameters ranging from 20 to 40 nm as a natural deviation in vesicle size population (46). Taking advantage of PLM sensitivity, our reported values are in agreement with previous reports of LUVs size distribution imaged via confocal microscopy and SEM (46), in addition to detecting small LUVs of $<r>$ $<50$ nm (Fig. S10).

**Membrane topography over NPs**

The demonstration of PLM performed here measured nanoscale hemispherical membrane curvature of an SLB draped over NPs ranging in radii from 24 to 70 nm. Prior methods of inducing nanoscale curvature utilized nanoengineered wavy glass substrates (47), microfabricated structures (48, 49), membrane tubule pulled from GUVs (50), and SLBs on deformable substrates (51, 52). However, wavy glass substrates, thick polymer structures, and lipid tubules are not compatible with TIRF excitation. The method of draping a membrane over fluorescent NPs of known size, as done here and previously (45, 53), was effective for engineering nanoscale membrane curvature, testing the capabilities of PLM, and revealing the effects of curvature on lipid mobility.

The NP sizes used here mimicked the sizes of membrane endosomes in cells. The top of the NPs was within the evanescent field and no fluorescence emission bleed from the NP was detected in the membrane channel. This was confirmed by multiple control experiments that are fully described in the supplemental material, and most clearly demonstrated here by the lack of localizations in sPLM at the site of the NP (Figs. 3 and 6). This comparison between sPLM and pPLM results provided confirmation of numerous aspects of our results. At the location of the NP-induced membrane curvature, a near uniform density of localizations in sPLM was detected whereas $> 5\times$ increase of localizations in pPLM was observed (Figs. 3 and 5A). This confirms that there was no significant chromatic bleed through from the fluorescence emission of the NP, that the refraction of the excitation light by the NP did not catastrophically alter the excitation light polarization, and that there was no significant Förster resonance energy transfer between DiI and the NP disrupted the polarization dependence of the signal. However, refraction by the NP may have influenced the direction of the DiI emission, as discussed below. When membrane curvature was created with LUVs rather than with NPs, there was an obvious increase in the number of localizations with sPLM as well as pPLM, as expected considering the difference in the membrane topography (Fig. S2).

The membrane topology over the NPs depended on the adhesion between the lipids and the polystyrene NP, the adhesion between the lipids and the glass coverslip, the size of the NP, the membrane bending rigidity, the lateral membrane tension, and the packing properties of lipids. We modeled the shape of the membrane over the NPs to be primarily spherical in shape with a smooth 20 nm radii of curvature bend to connect to the planar SLB (Fig. 5C). This consistent radius of curvature for the connection of the SLB on the NP to the SLB on the coverslip resulted in a tent-like transition from the top of the small NPs to the glass substrate and a neck-like feature at the transition from the large NPs to the coverslip. The tent-like membrane structure would have a bigger size than the small NPs; the neck-like membrane structure would have a smaller size than the large NPs. This trend is observed in the radial distributions of DiI localizations versus NP radius (Fig. 4). The tent-like model represents the initial stages of membrane bending upon the initiation of endocytosis or conclusion of exocytosis; the neck-like model represents the later stage of endocytosis or early stage of exocytosis.

The radial line scans of DiI localization density from pPLM over the NPs of $r_{NP}$ = 24, 51 and 70 nm as a function of distance from the center of the NP demonstrates sub-diffraction-limited resolution of membrane curvature (Figs. 4 and 5). The agreement between the experimentally measured and theoretically predicted radial density profiles suggest the accuracy of both the membrane model and the theoretical analysis of localization probabilities (Fig. 5 and Eq. 5).

**Limitations to resolution**

The distribution of localizations around the nanoparticle-induced membrane buds was influenced by multiple effects that limit the experimental determination of the membrane topography, including (1) localization imprecision of the individual fluorophores, (2) anisotropic emission from the membrane-confined DiI, (3) finite localization rates, (4) NP-induced emission lensing, (5) the fitting of multiple '*on*' fluorophores as if they were a single fluorophore, and (6) membrane curvature motion within the sample (*i.e.*, NP or LUV drift) (Fig. S9). Each of these contributions has been theoretically tested in attempts to match theoretical predictions to the experimental observations. It was found that the single-fluorophore localization imprecision and anisotropic emission proved to be the only error sources needed to theoretically reproduce the experimental data. Theoretical and experimental data matching required no NP-induced emission lensing nor multiple '*on*' fluorophore misassessments.

The inherent inability of DiI to tumble freely within the membrane is a necessary component for polarization sensitivity and membrane orientation detection; however, it also results in an anisotropic emission and systematic inaccuracies of DiI localization. While the DiI is not rigid in one location and can explore all $\psi$ values in addition to a tilt of $\beta = 69°$, some orientation averaging occurs where all these dipole orientations contribute to each single DiI image. Even still, the anisotropic emission results in a systematic shift up to 100 nm of the single fluorophore localizations towards the center of the NP.

SMLM is based on localizing single fluorophores that are sufficiently separated for computational fitting (>200 nm apart); however, if multiple fluorophores were in close proximity to each other (<100 nm) and falsely interpreted as a single fluorophore, then systematic errors could be incorporated. Typically, this error is predictable by assuming a uniform time-averaged fluorophore density, estimating the mean separation distance between fluorophores, and calculating the probability of multiple fluorophores being within the diffraction-limited range from each other. However, for pPLM, the assumption of a uniform time-averaged fluorophore density may not be appropriate. Since the horizontal membrane comprises the majority of the sample and DiI within the horizontal membrane absorbs less excitation light, it would follow that the sample-averaged fluorescence *off*-rate would be slower with p-polarized than with s-polarized excitation. This could yield more active fluorophores on the curved membrane than seemingly apparent. For the curved membranes examined here, if multiple fluorophore images were averaged simultaneously, the resulting localization will be shifted toward the center of the feature. The inclusion of this error caused worse fitting of our simulations to the experimental data, suggesting that the multiple '*on*' fluorophore misassessments were not a significant component of our image reconstruction and data analysis.

**Curvature affected lipid diffusion**

The sequential frame linking and analysis performed here resulted in the average trajectory sampling a distance of 180 nm, which is slightly less than half the circumference around the 70 nm radius NP. With greater experimental sampling densities, rates, and precision, a more sophisticated simulation and analysis routine would be warranted, the development of which is the focus of a later manuscript. However, the curvature-dependent step length observed here is dramatic and was able to be modeled computationally by incorporating the experimental data conditions, such as frame rate, localization precision, and anisotropic inaccuracies.

The diffusion of DiI apparently slowed when the membrane was curved over the NP. The change in membrane topography from flat to the curved membrane over the NP alters DiI diffusion observed in both s- and p-polarization, resulting in an 87% decrease in the observed diffusion coefficient through the $xy$-plane ($D_{xy}$) (Fig. 7). Since the diffusion analysis from the sPLM and pPLM data yielded indistinguishable effects of membrane curvature on lipid mobility, the illumination polarization did not affect the observed diffusion coefficients. When a membrane is tilted ($\theta > 0$), a 2D Brownian diffuser apparently moves slower as imaged in the $xy$-plane; however, this geometrical effect alone was not sufficient to reproduce diffusion rates extracted from experimental data. Further, a diffusion barrier between the membrane bud and the surrounding planar SLB was not sufficient in matching the modeled and experimental data. However, combining both the geometrical effects of the tilted membrane and a curvature-dependent membrane viscosity yielded a strong agreement between the modeled and experimental SPT results. This analysis supports the hypothesis that DiI diffuses slower on more curved membranes due to change in membrane properties such as membrane viscosity or lipid packing, as suggested previously (45).

Neither the experimental data nor the simulated theoretical reproduced data for $D_{xy}$ distinguishes between the leaflets of the lipid bilayer. The SLBs were symmetrically labeled through the addition of DiI to the lipid mixture before GUV electroformation, and presumably, both bilayer leaflets contributed to the observed DiI diffusion rates. DiI in the outer leaflet would have with minimal direct substrate interaction, whereas DiI in the inner leaflet would be in close proximity to the polystyrene NP or glass support. The slowing of the molecules at membrane curvature regions has also been reported before where the diffusion rate exhibit confined slow diffusion in comparison to flat regions (53).

**Future improvements to PLM**

PLM is able to provide super-resolution detail on membrane orientation with improved sensitivity and resolution from comparable methods. Since PLM requires no manipulation of the fluorescence emission path or the Airy function PSF, the incorporation of PLM with SMLM in additional complementary color channels is straightforward. The simultaneous super-resolution membrane orientation detection via PLM with the curvature sorting and induction effects of cholera toxin subunit B (CTxB) is the focus of a companion manuscript in this journal [cite CTxB paper].

It is feasible that the local membrane orientation could be evaluated by the direct mapping of acquired localizations per pixel to the PLM theory. In order to perform such analysis, a minimal localization density of >0.05 localizations/nm$^2$ would be required. PLM has the advantage of observing lipids that diffuse into the region of view from the surrounding membrane to effectively achieve unlimited labeling densities, similar to as has been previously utilized in point accumulation for imaging in nanoscale topography (PAINT) (54). Greater sampling statistics would enable finer details of membrane topology to be extracted with more statistically significant comparisons between pPLM and sPLM localization densities.

With a faster frame rate and/or decreased localization imprecision, more sophisticated SPT analysis could be performed. It would be instructive to analyze the single-molecule trajectories in such a way to extract the component of the molecular diffusion that changes the molecule's distance from the center of the bud (*i.e.,* radial distance) as opposed to the diffusion component around the bud (*i.e.,* azimuthal distance). If some membrane components accumulate at the bud neck, it is feasible that the single-molecules could diffuse quickly around the bud while slowly changing its radial distance from the bud center. Unfortunately, this analysis is not feasible at the current imaging frame rates, but it will be the focus of future work.

**CONCLUSIONS**

Polarized localization microscopy (PLM) is capable of detecting and resolving nanoscale membrane curvature with super-resolution and correlating this curvature to the single-molecule diffusion and molecular sorting. PLM requires no alteration of the emission path from traditional single-molecule fluorescence microscopes and incorporates no inherent sacrifice in the signal or localization precision for observing the membrane orientation. Distinct identification between membrane topology of LUVs, unfused GUVs on SLBs, and curved membranes over NPs were observed. The nanoparticle-patterned substrate provided a suitable platform to engineering nanoscale membrane curvature of physiologically relevant dimensions. Local membrane bending regions with radii of curvature $\geq 24$ nm was detected. PLM detects membrane curvature and resolves membrane topography as soon as within 1 sec of acquisition time with $(1.2 \pm 0.1) \times 10^{-6}$ localizations/nm$^2$/frame.

      Radial line scans of pPLM localizations reveal radii of curvature of $32 \pm 4$, $50 \pm 14$, $60 \pm 13$ nm for membranes over the nanoparticles radii of 24, 51, and 70 nm, respectively. Further, a 6x increase in the SNR is obtained by PLM over traditional TIRFM. The theoretically estimated localization probabilities versus membrane orientation well reproduced experimental data. The unique spatiotemporal resolution of PLM is suited to monitor membrane structure variation with lipid and protein dynamics. We envision that this microscopy technique will aid in providing new information for previously untestable nanoscale processes coupled with a change in membrane topography. This was demonstrated by the observation of time-dependent membrane budding initiation and growth induced by cholera toxin subunit B in quasi-one component lipid bilayers, revealing a possible mechanism of cholera immobilization and cellular internalization described further in the accompanying manuscript [cite CTxB]. Fundamental questions regarding nanoscale cellular processes including clathrin-independent endocytosis, viral infections, endocytosis/exocytosis, and immunological responses are soon to be addressed with PLM. The feasibility of performing PLM on model membranes or live cells on time scales suitable for observing cellular processes permits this technique to be adopted and broadly used to probe cellular dynamics.

## ACKNOWLEDGEMENTS

The authors thank Arun Anantharam, Dipanwita De, Henry Edelman, Rebecca Meerschaert, and Eric Stimpson for valuable discussions. A.M.K. was funded by Thomas C. Rumble Fellowship Award. Financial support was provided by Wayne State University

## AUTHOR CONTRIBUTIONS

A.M.K. and C.V.K. designed the experiments, analyzed the data, and prepared the manuscript. A.M.K. performed the experiments.

## COMPETING FINANCIAL INTERESTS STATEMENT

The authors have no competing financial interests.

**FIGURES**

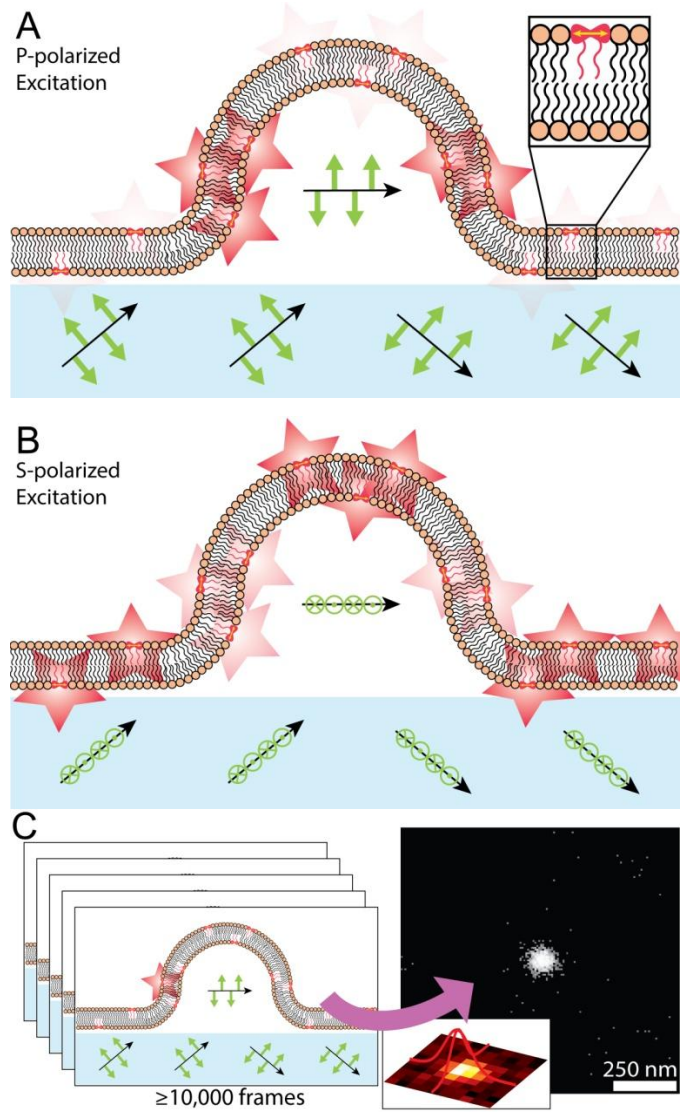

**FIGURE 1** Polarized Localization Microscopy (PLM) combines the techniques of polarized TIRFM and SMLM. By controlling the linear polarization of incident excitation light, the electric field (*green arrows*) of the evanescent wave for fluorescence excitation can be either (A) vertical with p-polarized light or (B) horizontal with s-polarized light. This results in differential fluorophore excitation depending on membrane orientation. (C) Imaging and localizing individual blinking fluorophores in separate frames enable the reconstruction of super-resolution images with embedded information on membrane orientations.

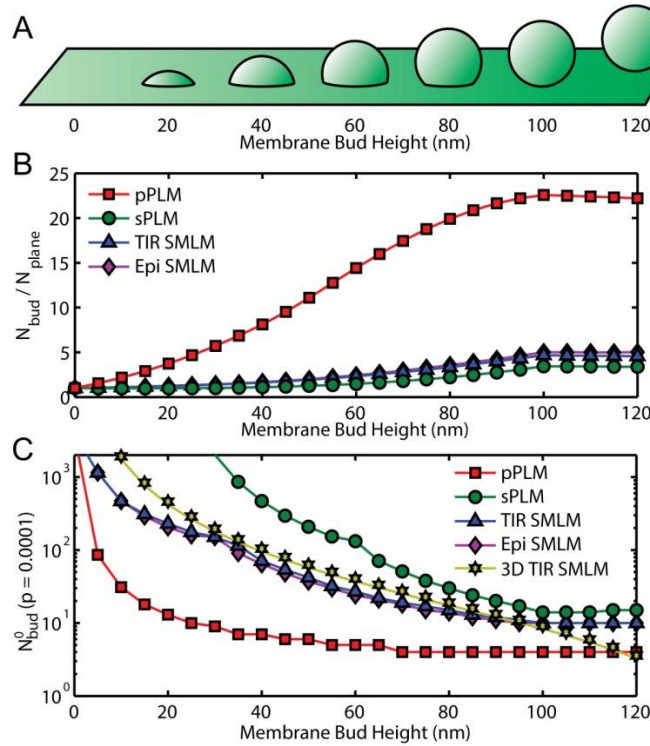

**FIGURE 2** Estimation for observed localizations of a budding membrane reveals the sensitivity of PLM compared to other optical methods. (A) Membranes containing buds of 50 nm radii of curvature were analyzed at varying protrusion distances from the surrounding connected planar membrane. The fractional increase in localizations due to the bud are plotted. (B) The increased number of localizations expected due to non-planar membrane shape relative to the number of localizations expected from a planar membrane demonstrated a 22.5x increase in localization density with pPLM, which is over 4x larger than expected for sPLM, unpolarized epifluorescence SMLM, and unpolarized total internal reflection (TIR) illumination SMLM. (C) The required number of localizations to identify a membrane bud from the surrounding SLB ( $N^0_{bud}$ ) with $p = 0.0001$ are plotted for pPLM, sPLM, unpolarized TIR, and unpolarized epifluorescence SMLM from Eq. 7 and for 3D SMLM from Eq. 8.

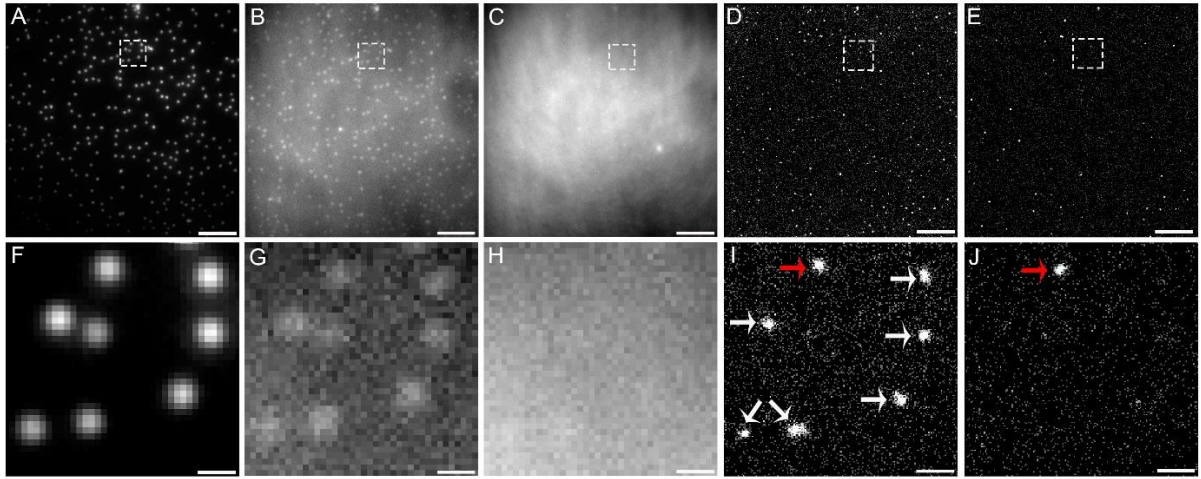

**FIGURE 3** Membrane curvature was engineered by draping a supported lipid bilayer over NPs. (A) The 70 nm radius fluorescent NPs on glass were imaged with $\lambda_{ex}$ = 488 nm. (B-C) Diffraction-limited p-polarized and s-polarized TIRFM image, respectively. (D-E) Reconstructed images of the membrane over the NPs presented as 2D histograms of the localizations in pPLM and sPLM, respectively. (B-D) The membrane was imaged with $\lambda_{ex}$ = 561 nm and the differences between the polarizations provide internal controls for chromatic bleed through. (F-J) Magnified images from (A-E), as indicated by the dashed boxes. The detected membrane curvature over the 70 nm NPs is indicated by white arrows. A multi-colored fiduciary mark is indicated by red arrows. Scale bars represent (A-E) 5µm, and (F-J) 400 nm. Similar results for LUVs and other sizes and colors of NPs are shown in the supplemental information.

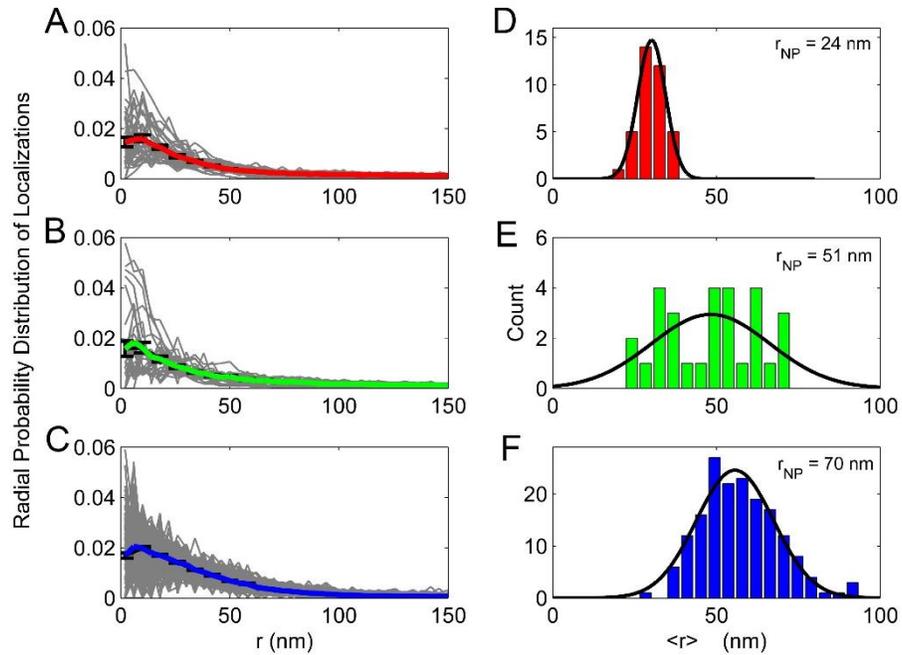

**FIGURE 4**: (A-C) Radial probability distribution of localizations versus distance from the center of the curved membrane (*r*) over NPs for $r_{NP}$ = 24, 51, and 70 nm, respectively. Grey lines represent individual events and the colored lines represent the average. Error bars are the standard error of the mean at a given *r*. (D-F) Histograms of the radius for each curvature event (<*r*>) over NPs of $r_{NP}$ = 24, 51, and 70 nm, respectively. Black lines represent the Gaussian fits to guide the eye. The mean of the events radii was 32 ± 4, 50 ± 14, 60 ± 13 nm for $r_{NP}$ = 24, 51, and 70 nm, respectively.

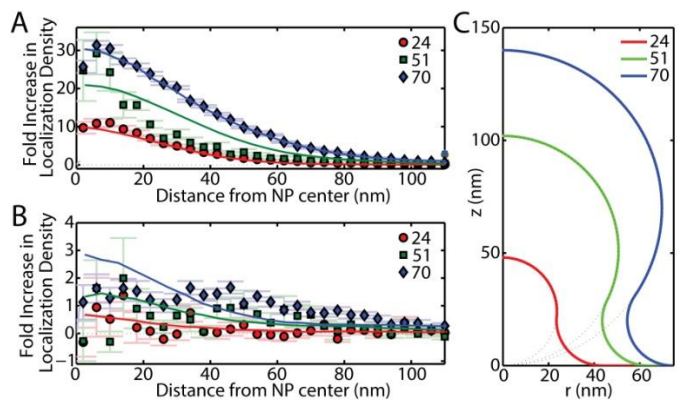

**FIGURE 5** Radial density line scans for localizations from membrane topography observed in (C) over $r_{NP}$ = 24, 51, and 70 nm via (A) pPLM and (B) sPLM. Data points show the experimental results while the simulation results are plotted as solid lines. A single $B_0$ value and out-of-focus magnitude for anisotropic inaccuracy were fit all six data sets.

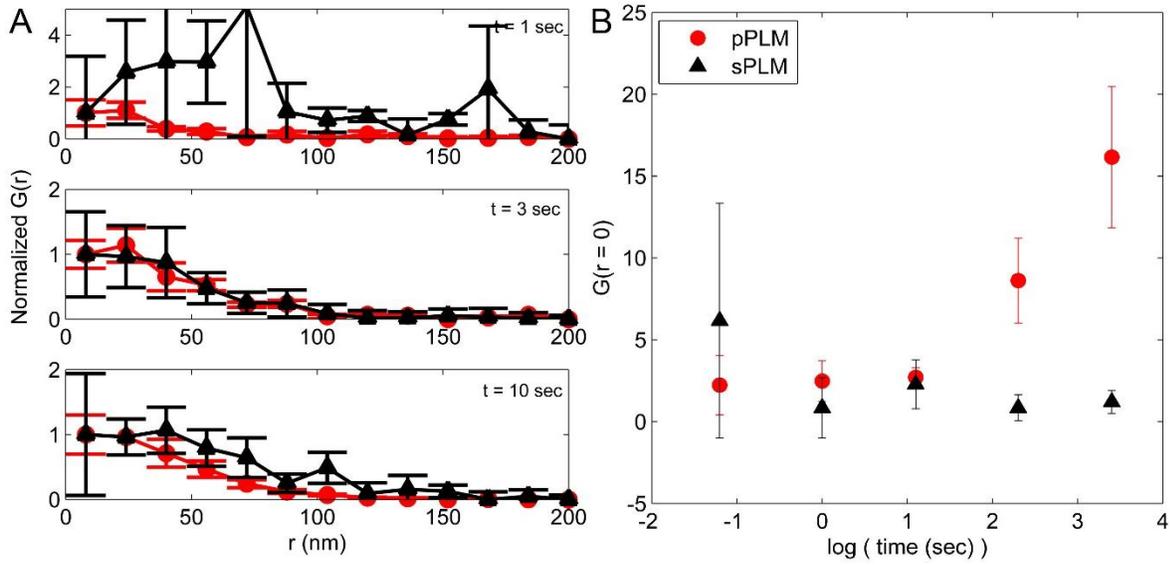

**FIGURE 6** Autocorrelation analysis for sPLM (black) and pPLM (red) for different acquisition time intervals. (A) Data was normalized to the first G(r) value to show the detection of curvature in pPLM with smaller error bars as time interval increases in comparison to sPLM. (B) Autocorrelation G(r) value at $r = 0$. The sPLM analysis provides the internal control to show the uniform bilayer analysis. An acquisition time of 1 sec is sufficient to indicate the presence of curvature in pPLM.

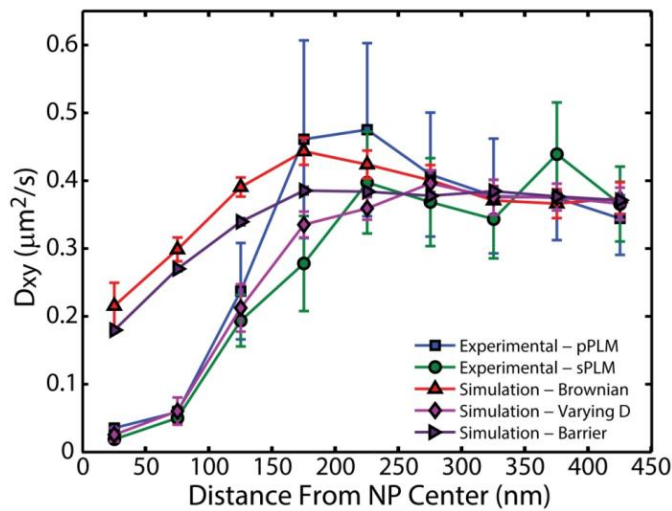

**FIGURE 7** Single-particle tracking of DiI molecules reveals slowed diffusion at the site of nanoscale membrane curvature equally while imaged with p- or s-polarized excitation. SLBs were draped over 70 nm radius NPs. Particle locations were projected on the *xy*-plane and the apparent fluorophore diffusion was affected by both the 3D membrane topology and the influences of membrane curvature on DiI mobility. The fit of the distribution of step lengths to Eq. 1 yielded the apparent diffusion coefficient and the 95% confidence range, as indicated by the error bars. Neither a locally Brownian diffusion nor a simulated barrier to free diffusion surrounding the bud reproduced the experimental results. However, upon simulating a decreased local *D* for curved SLB to 13% of the planar SLB value, the resulting simulated $D_{xy}$ well matched experimentally observed $D_{xy}$.